\documentclass[12pt, onecolumn, draftclsnofoot]{IEEEtran}
\usepackage{amsmath,epsfig}
\usepackage{amssymb}
\usepackage{amsfonts}
\usepackage{array,dcolumn}
\usepackage[ruled,vlined,linesnumbered]{algorithm2e}
\usepackage{psfrag}
\usepackage{amsmath}
\usepackage{amsfonts}
\usepackage{amssymb}
\usepackage{accents}
\usepackage{booktabs, adjustbox}
\usepackage{cite}
\usepackage[all]{xy}
\usepackage{dsfont}
\usepackage{url}
\usepackage{xcolor}
\usepackage[nolist]{acronym}
\usepackage{graphicx}
\usepackage{subfig}
\usepackage{threeparttable}

\title{Communication Aspects of the Integration of Wireless IoT Devices with Distributed Ledger Technology}
\author{\IEEEauthorblockN{Pietro Danzi\IEEEauthorrefmark{1}, Anders E. Kal{\o}r\IEEEauthorrefmark{1}, Ren\'{e} B. S{\o}rensen\IEEEauthorrefmark{1}, Alexander K. Hagelskj{\ae}r\IEEEauthorrefmark{2}, \\ Lam D. Nguyen\IEEEauthorrefmark{1}, \v{C}edomir Stefanovi\'c\IEEEauthorrefmark{1} and Petar Popovski\IEEEauthorrefmark{1}}\\
\IEEEauthorblockA{Department of Electronic Systems, Aalborg University, Denmark}\\
\IEEEauthorblockA{E-mail: \IEEEauthorrefmark{1}\{pid, aek, rbs, ndl, cs, petarp\}@es.aau.dk, \IEEEauthorrefmark{2}ahagel14@student.aau.dk}
}

\begin{document}
\maketitle

\begin{abstract}
The pervasive need to safely share and store information between devices calls for the replacement of centralized trust architectures with the decentralized ones.
Distributed Ledger Technologies (DLTs) are seen as the most promising enabler of decentralized trust, but they still lack technological maturity and their successful adoption depends on the understanding of the fundamental design trade-offs and their reflection in the actual technical design.
This work focuses on the challenges and potential solutions for an effective integration of DLTs in the context of Internet-of-Things (IoT).
We first introduce the landscape of IoT applications and discuss the limitations and opportunities offered by DLTs.
Then, we review the technical challenges encountered in the integration of resource-constrained devices with distributed trust networks.
We describe the common traits of lightweight synchronization protocols, and propose a novel classification, rooted in the IoT perspective.
We identify the need of receiving ledger information at the endpoint devices, implying a two-way data exchange that contrasts with the conventional uplink-oriented communication technologies intended for IoT systems.
\end{abstract}

\section{Introduction}

In recent years, there has been an enormous interest in applications of Distributed Ledger Technologies (DLTs) to the realm of Internet of Things (IoT).
The motivation stems from the lack of a pervasive IoT trust layer: IoT applications are usually confined within their own ecosystem, making it difficult to share authenticated information among different applications.
DLT is seen as an enabler of a distributed trust network, that would be used by IoT devices to manage and exchange data without involving centralized control~\cite{ali2018applications}.
This would not only replace some of the existing centralized applications with their decentralized versions, but also lead to a wave of completely new applications, such as trusted access to edge computing resources~\cite{yang2019integrated}.

However, the task of meaningful integration of DLTs into existing networking infrastructures calls for an investigation of the challenges posed by DLTs~\cite{neudecker2018network}.
The key problems are the cost of storing the ledger and verifying its correctness, as well as the scaling of the frequency of the ledger updates.
In addition, an insufficient attention has been dedicated to issues related to the integration of DLT into wireless IoT systems~\cite{danzi2018analysis}.

This paper addresses the following question: \emph{What is the capability of the wireless IoT technologies available on the market for supporting the traffic generated by DLTs?}
In order to provide the answer, we first classify IoT applications and map them into DLT-aided applications, which is the basis to understand and quantify the communication demands of DLTs.
After identifying the pivotal role that lightweight synchronization protocols play in the integration of DLTs into wireless IoT systems, we study two practical use cases: (1) Integration of DLT with a measurement reporting system based on LoraWAN; 
(2) Integration of DLT into an IoT application based on the consensus among the agents, e.g., in a distributed control system.
These simple, but illustrative cases show that two-way wireless communication is required to enable high decentralization.
Another conclusion is that the achievable level of decentralization is largely determined by the wireless technology used.

The rest of the text is organized as follows.
Section II provides background on DLTs and IoT services.
Section III arguments the challenges of integrating IoT with DLTs.
Section IV presents the architecture of a distributed trust network.
Section V introduces and classifies the lightweight synchronization protocols.
Section VI elaborates two case studies.
The paper is concluded in Section VII.

\section{Background on DLTs and IoT Services}

\subsection{Distributed Ledger Technologies (DLTs)}

Distributed ledger technology involves a set of protocols designed to replicate a timestamped and ordered database, termed \emph{ledger}~\cite{nakamoto2008bitcoin}.
The nodes of a DLT network store a copy of a common ledger, that is kept consistent by means of hash chaining.
To append new data, termed \emph{transaction}, to the ledger, the nodes should fulfill a set of rules defined by a consensus mechanism, e.g., based on proof-of-work~\cite{nakamoto2008bitcoin}.
The aim of consensus mechanisms is to ensure consistency of local copies, while limiting the ledger update rate.
For instance, in blockchains, a bulk of valid transactions can occur from every few seconds (as in Ethereum) to minutes (in Bitcoin).
In DLTs based on directed acyclic graphs, e.g. IOTA, a computation has to be done to update the ledger.
It was experimentally shown that this computation requires several seconds for low-energy devices~\cite{elsts2018distributed}.
Moreover, in DLTs, an already included transaction may be reverted (i.e. removed from the ledger), if a consensus is reached on a different and conflicting transaction.
Hence, DLTs also introduce the concept of ``finalized transaction", referring to an accepted transaction that will not be reverted with a high probability.
The finalization probability is an increasing function of the time since the transaction was published.
This introduces a trade-off between the finalization certainty and the tolerated transaction validation delay~\cite{danzi2018delay}, depending not only on the application, but also on the consensus mechanism of the specific DLT.

\subsection{IoT Devices and Services}

As envisioned by 5G standardization~\cite{one5g}, IoT services can be divided into two general categories: massive Machine Type Communication (mMTC, also known as massive IoT) and ultra-reliable low-latency (URLLC).
The main distinctions between them are seen in the requirements on communication latency, reliability and scalability, but also in the capabilities of the devices.

A typical mMTC application involves a sensor network in which devices report their observations to a server in quasi-periodic manner, with a low factor of activity~\cite{one5g}.
The main challenge here is the massive number of reporting devices; 5G mMTC use-cases foresee from tens of thousands to million connections per 1~$\text{km}^2$~\cite{one5g}.
On the other hand, reliability and latency requirements in mMTC are significantly less challenging compared to URLLC.
For instance, the target reliability and latency of user-plane radio connections for applications like smart city or factory monitoring are 95\% and 0.5 s, respectively~\cite{one5g}.

URLLC applications feature general requirements of user-plane radio reliability of 0.99999 and latency of 1~ms~\cite{one5g}.
Traffic patterns depend on the pplication and can be deterministic periodic, quasi-periodic and event-driven reporting.
Deterministic periodic traffic is characteristic for real-time control applications, which operate in cycles of the order of 1~ms and have extreme requirements, e.g. user-plane radio reliability of 0.999999999 and latency of 0.1 ms in industrial automation~\cite{one5g}.
Event-driven reporting is characteristic for alarm systems, where some event(s), e.g. detection of a fire, triggers the devices.
An additional challenge here is a high spatio-temporal correlation of the report generation among the triggered devices, considerably straining the wireless network.

In any IoT application, every device acts as an information publisher and/or subscriber.
As publisher, the device sends its data to servers, e.g., MQTT brokers.
Depending on the application, the device may also require the receipt of the data publication.
When acting as a subscriber, the device informs the servers about its topics of interest (usually, in the initialization phase) and the servers push the information to the device, when available.
For the introduced categories of IoT applications, an example of publisher is a sensor in mMTC; a subscriber could be an actuator that takes action after an alarm is reported.
Finally, the devices in URLLC applications are, generally, more capable than the mMTC devices in terms of computational and power supply.
Thus, URLLC and mMTC devices face different challenges in the integration of DLTs: while the transaction finalization delay poses a problem for URLLC devices, mMTC devices are challenged by the computational requirements and the protocol overhead.

\section{Will DLTs fit with IoT applications?}

In this section, we discuss the challenges of integrating DLTs in IoT ecosystem, focusing on the communication aspects.
We consider solutions in which the IoT devices do not store the ledger, hence, the continuous growth of the ledger size has a little impact on the storage costs of IoT devices.
However, it still involves the transmission of a high amount of the metadata, e.g., the block headers in blockchains, contrasting the standard perception that the IoT traffic has very small payloads.
A related issue is that distributed trust architecture symmetrizes the uplink and downlink traffic, conflicting with the fact that a number of mMTC technologies have been designed and optimized for uplink-dominated traffic~\cite{danzi2018delay}. 

A second obstacle is seen in the latency of the publishing of information to the ledger, which adds up to the latency of the communication system.
Specifically, the validation delay prevents the existing DLTs in supporting real-time and event-driven reporting.
Together with the finalization delay, this prevents the use of DLTs for URLLC applications.\footnote{E.g., in Bitcoin network, the validation usually takes ten minutes, and a transaction is considered finalized with high probability after tens of minutes. An approach to overcome this problem are the off-chain payment channels~\cite{hannon2018bitcoin}.}
We quantify verification and finalization delays for several DLTs in Section~VI.

Another challenge is that the ledger replications make the cost of operating IoT servers high, since these have to constantly store and verify the ledger updates.
This harms the network decentralization, as a large number of IoT devices will be connected to relatively few servers.
Secondly, the overhead due to metadata (e.g. digital signatures) makes storing in DLTs the large amounts of data, commonly required by quasi-periodic reporting applications, inefficient.
Thus, the devices are motivated to limit the amount of information stored in the ledger by being required to pay fee to the DLT node that validates their transaction (in blockchains) or to spend energy (in IOTA) when creating a transaction.

Finally, DLT systems do not provide service differentiation, which poses challenges for IoT applications that need high reliability guarantees.
Instead, DLTs are usually application-agnostic~\cite{androulaki2018hyperledger} and the transaction prioritization is delegated to the endpoint devices, which indicate to the DLT network the importance of the data they are transmitting, e.g. by paying a fee.
Hence, the validation delay of a transaction depends on its priority w.r.t. other transactions waiting to be validated by the DLT network.

Due to all these challenges, DLTs may not be suitable for IoT applications that require short delays, or involve very simple devices and/or networks with limited downlink capabilities.
However, DLTs may still be suitable for complementing a vast class of IoT applications with a sporadic interaction between devices and DLTs; e.g. accountability and access management to resources, such as edge computing~\cite{yang2019integrated}.
The ideal DLT use-case is the one that has the availability of trusted data as the most important feature and can tolerate a data finalization delay.

In conclusion, the benefit of integrating a DLT with existing applications lies in the formation of a new distributed trust ``layer".
The architecture of such distributed trust network is discussed next.

\section{The architecture of Distributed Trust Networks}

\begin{figure}[t]
\centering
  \subfloat[]{\includegraphics[width=0.75\columnwidth]{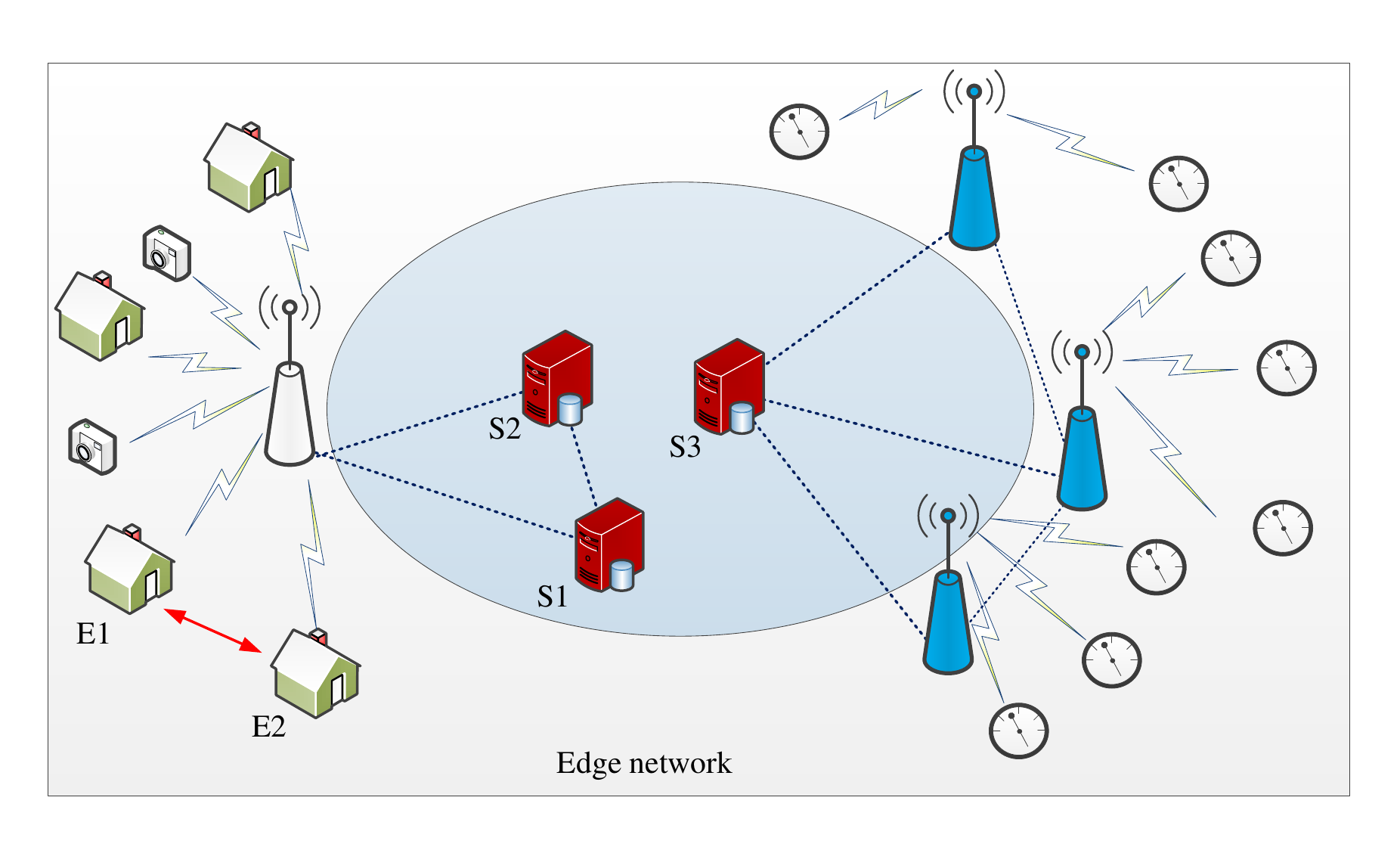}}
  
   \subfloat[]{\includegraphics[width=0.75\columnwidth]{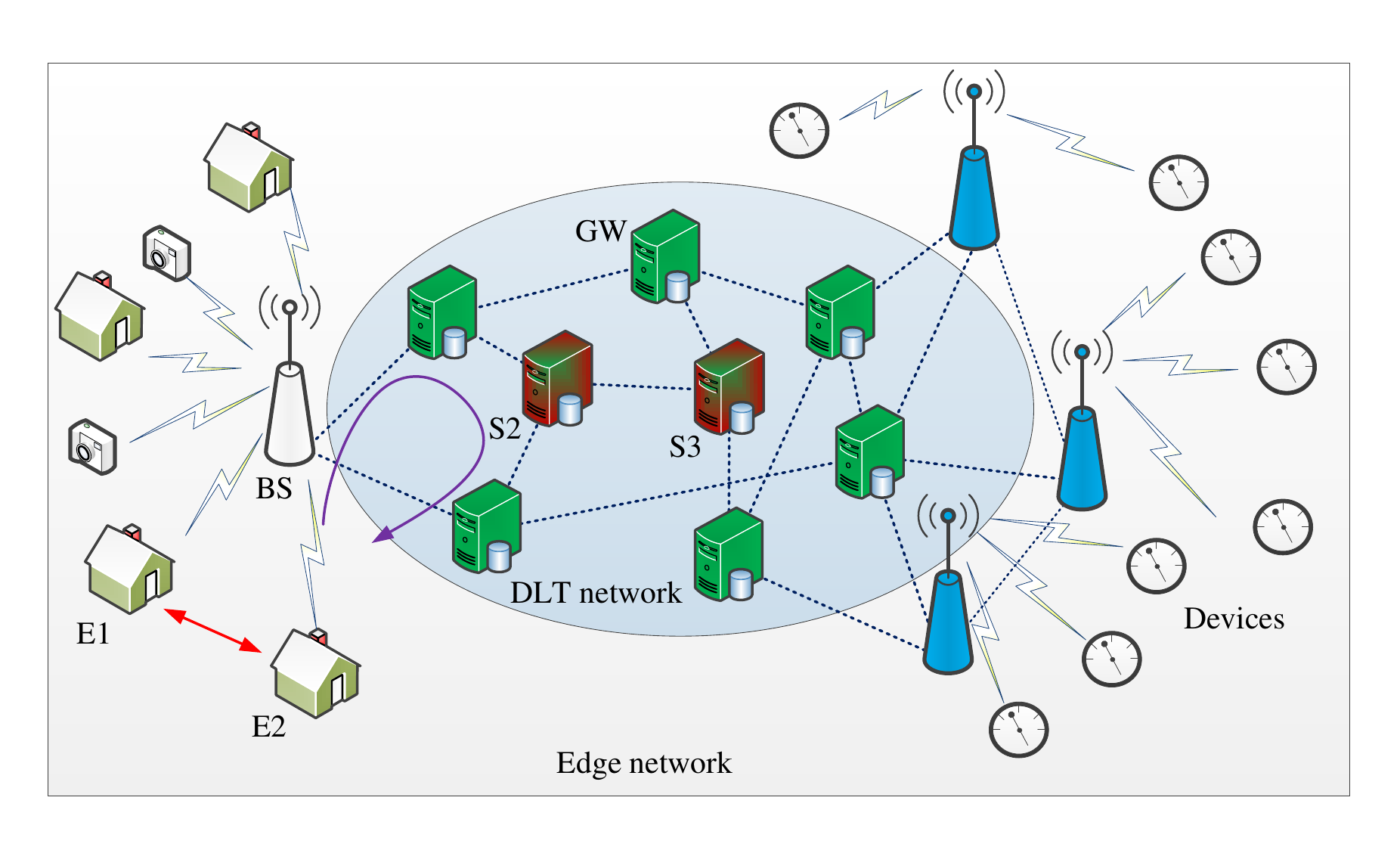}}
    \caption{The (a) legacy system architecture and (b) DLT-based. S1, S2 and S3 are servers, E1 and E2 are endpoint devices. In (b), the green nodes are gateways (GWs), while S2 and S3 are both servers and GWs. Dotted lines represent network links.}
    \label{fig:architecture}
\end{figure}

A typical architecture of IoT applications is depicted in Fig. 1(a): devices interact with a server that provides secure and reliable write/read access to a database.
The fundamental issue of such architecture, motivating the introduction of DLTs, is a low interoperability among servers.
For example, two endpoint devices (E1 and E2 in Fig.~1) can exchange authenticated information only if their servers support this functionality; this is not the case for S1 and S3 in Fig.~1(a). 

The transition to a DLT-based distributed trust network requires changes in the IoT networking architecture, as depicted in Fig. 1(b). Here, the servers and the edge devices are interconnected by the DLT network, which allows sharing of data authenticated by the DLT.
The DLT networks usually rely on wired technologies and consist of nodes that verify and store the updates to the ledger.
Some of these serve as gateways (GWs) between the IoT devices and the DLT.
The interactions between GWs and IoT devices are typically regulated by lightweight synchronization protocols, designed to simplify the interaction with the distributed trust network.
However, as presented in the first case study in Sec.~VII, some implementations of such protocols are unsuitable for LoraWAN networks, especially in regards to delivering blockchain block headers to a large number of wireless devices.
Another issue is the low energy available to certain IoT devices, which prevents them from initiating frequent interactions with GWs. For this reason, there exist several types of lightweight synchronization protocols that procure different costs and benefits. These are categorized and discussed in the following section.

\section{Protocols for Distributed Trust Networks}

\subsection{Lightweight Synchronization Protocols}

Protocols that facilitate lightweight ledger synchronization have been studied in~\cite{danzi2018delay}, where three protocol classes P1, P2, and P3, with different trust levels were identified. Protocols from class P1 provide the device with all information necessary to verify the DLT, %
 hence, the amount of trust outsourced the network is minimal.
Such protocols are not envisaged for implementation in IoT devices due to high memory and communication requirements~\cite{danzi2018analysis}.
Class P2 represents protocols in which the IoT devices receive a reduced amount of ledger information, only being capable of verifying that the received information is consistent, but not whether the transactions are valid.
For instance, in Bitcoin Simplified Payment Verification (SPV) protocol~\cite{nakamoto2008bitcoin}, a client only receives the transactions in which it is interested, together with a reduced amount of metadata, i.e., block headers and Merkle proofs.
The client is thus incapable of verifying the validity of the transaction, and can only verify whether the metadata received from different GWs is the same.
Finally, P3 protocols delegate the full trust to the GWs and, hence, are not attractive for distributed trust networks.
However, they might be the only viable solution in certain IoT systems, as further elaborated in Sec.~VI.

In the context of distributed trust networks, for which the class P2 is the only suitable one, it is helpful to further distinguish between two subclasses:
\begin{enumerate}
    \item \emph{Digest-based protocols}, in which the GWs push to the client a digest (e.g. block header) of the information appended to the ledger. Based on the digest, the client may request further information of interest, e.g. a transaction validated in the past. This class of protocols stems from SPV protocol.
    \item \emph{Incentive-based protocols}, in which the GWs push the information of interest to the client, when available. Since the client lacks means to verify the trustworthiness of GWs, deposits of a certain value are set up to ensure that the GWs act honestly. To increase the security, further mechanisms are implemented, such as charging fines to GWs that send false data~\cite{gruber2018unifying}.
\end{enumerate}

On top of the DLT synchronization protocols, IoT devices can run decentralized applications (``dApps")~\cite{ali2018applications,christidis2016blockchains,danzi2017distributed} or execute ``off-chain" protocols, that allow the exchange of temporary information among devices~\cite{hannon2018bitcoin}.
In the simplest formulation of an off-chain protocol, two devices initiate a payment channel on the ledger by each sending a transaction to the ledger.
Once the transactions are validated, they exchange data without involving the DLT network.
To close the channel, each device sends a second transaction confirming that the previously exchanged data is trusted by both parts.

\subsection{Protocol Overhead}

\begin{figure}[t]
\begin{center}
\includegraphics[width=0.5\columnwidth]{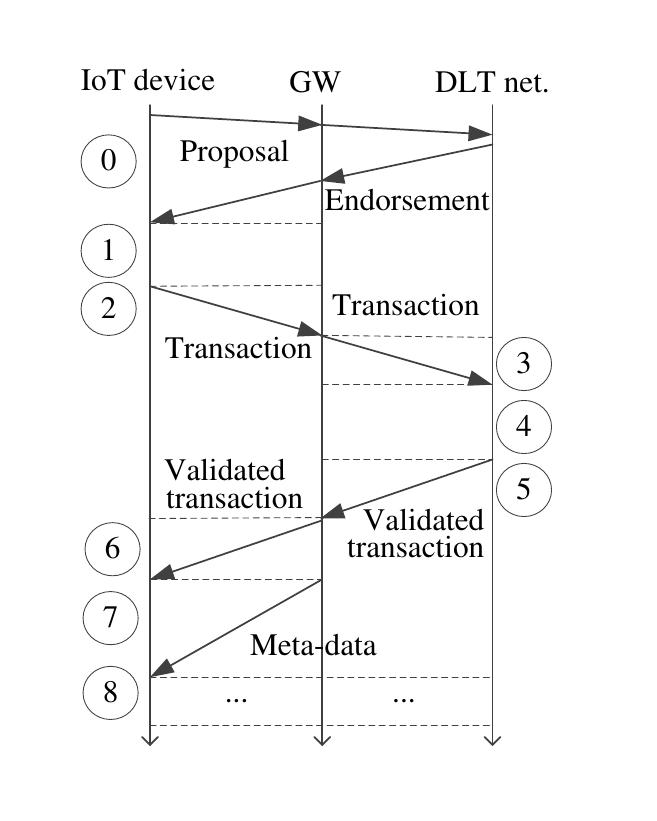}
\end{center}
\caption{A sequence of message exchange in DLTs. The phases are numbered from 0 to 8.}\label{fig:messages}
\end{figure}

We now return to the discussion of the protocol overhead imposed by DLTs.
A general message exchange in lightweight protocols is depicted in Fig.~2; note that some specific protocols implementations, e.g. Bitcoin~\cite{nakamoto2008bitcoin}, Ethereum, Fabric~\cite{androulaki2018hyperledger} and IOTA, skip some of the phases.
For instance, phase 0 exists only in Fabric, serving to prepare the transaction by collecting the ``endorsement", i.e. authentication, from nodes of DLT network via the GWs.
In phase 1, the IoT device invests energy to solve the PoW (this is only encountered in IOTA) and digitally signs the transaction (this happens in all the protocols); then, in phase 2, it sends the transaction to the GW.
The transaction is forwarded to the DLT network in phase 3, and included in the ledger at the end of phase 4 (this delay exists in blockchains but not in IOTA, where the PoW was done already in phase 1).
In phase 5, the GW is informed of the validation of the transaction, subsequently informing the IoT device in phases 6 and 7.
For instance, in Ethereum-like blockchains, the message in phase 6 is a receipt and the one in phase 7 contains block headers and the proof of inclusion~\cite{danzi2018delay}.
In incentive-based protocols, step 6 may not be present, and the message in phase 7 may just contain an acknowledgement from a trusted GW.
Observe that with incentive-based protocols, the amount of DLT metadata sent to the device is minimal, because it is filtered by the GWs~\cite{gruber2018unifying}.
This is not the case for digest-based protocols, which feature a large metadata overhead to ensure authenticity and reliability of the information.

The overhead related to authenticity is due to the fact that GWs cannot be entirely trusted.
In the uplink, the transaction is, when possible, locally signed by the device before being transmitted.\footnote{Previous works showed that the operation of digital signature may be too expensive for certain classes of IoT devices~\cite{elsts2018distributed}.}
In the downlink, by receiving (part of) the ledger information, a device can verify that the GWs are effectively forwarding its transactions to the DLT network.
The issue is that the device must trust that the GWs are timely forwarding correct ledger information.
The detection of malicious GWs in lightweight protocols is usually handled by establishing connections to multiple GWs to reduce the possibility of collusion.\footnote{Discovery of new GWs in a decentralized fashion introduces additional overhead that is out of the paper scope. The limited resources available at IoT devices may impede this process.}
For instance, in Fig. 1(b), E2 sends a transaction via a GW and receives the feedback from a second one (denoted by the purple arrow).
Another advantage of connecting to multiple GWs lies in the possibility to hide the sources, that a device subscribes to, by subscribing to different sources at each GW.
Finally, devices participating in off-chain payment channels can check if the channels are correctly opened/closed.

A feedback channel is also required in order to achieve reliability.
First, in blockchains, the devices need information regarding the priority given to their transaction w.r.t. the state of the network.
In other words, they need to be informed about the transaction fee to pay to the network. Another motivation for feedback is that a transaction may fail due to conflicts with transactions from other devices.
This may happen, e.g. in Ethereum smart contracts, where portions of general-purpose memory stored in the DLT can be shared between several devices.

\section{Integration of DLTs with wireless IoT technologies}
\begin{figure}[t]
\begin{center}
\includegraphics[width=0.85\columnwidth]{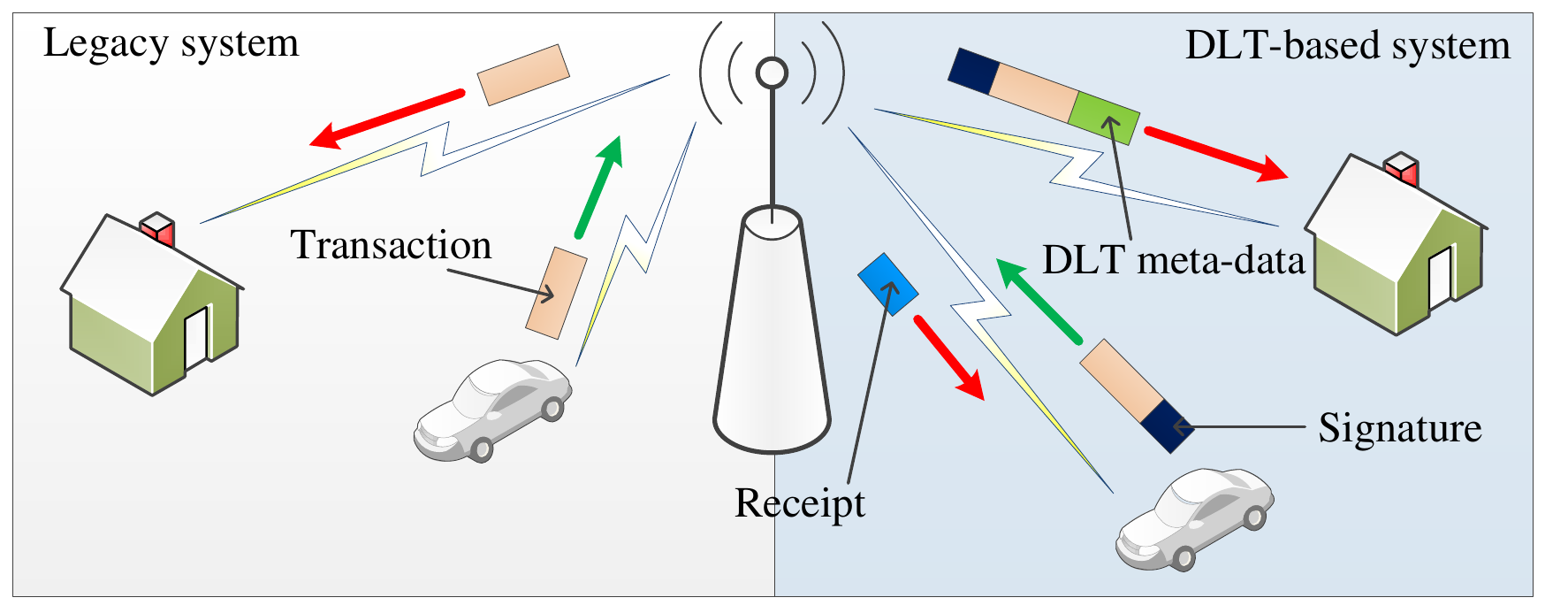}
\end{center}
\caption{Information sent within the wireless access point and the publishers (cars) and subscribers (houses), for legacy and DLT-based systems.}\label{fig:pubsub}
\end{figure}

The lightweight protocols introduce a number of additional steps compared to a system with centralized trust.
In the publish operation, depicted in Fig. 3, there is the additional burden of digitally signing the transaction and of transmitting the signature.
Depending on the application, the device may also require a receipt that allows it to verify that the information was successfully published.
Furthermore, a device acting as a subscriber needs to receive not only the information published to the subscribed channel, but also the publisher's signature and the metadata that allows the device to verify that the information can be trusted (i.e. that it is included in the ledger).
Both the receipt sent to the publisher and the metadata sent to the subscriber to verify the published information impose protocol overhead, which decreases with the amount of trust that the device outsources to the network.

From the perspective of the wireless access network, the main difference between distributed trust networks and legacy architectures lies in the involvement of IoT devices in new protocols for distributed trust. %
In the following subsections, we demonstrate some of the issues encountered when using popular DLTs in LoraWAN and IEEE 802.11 wireless networks, which are widely used for IoT. 

\subsection{Case study 1: Publishing and subscribing to a DLT via LoRaWAN}

LoRaWAN is a low-power wide area network technology %
which enables long-range communication at modest data rates.
In Europe, LoRaWAN operates in ISM bands, where is imposed a duty-cycle of 1\%, which is enforced by waiting for 99 transmission times after each transmission.
This constraint limits the amount of downlink traffic from the access point (AP), challenging the suitability of LoRaWAN for devices that subscribe to information streams.
We shall investigate subscribing and publishing in the following paragraphs.

\begin{table} 
\caption{Parameters of the DLTs considered in the case study.}
\label{tab:dlts}
\centering
\begin{tabular}{p{1cm}p{0.13 \linewidth}p{0.13\linewidth}p{0.13\linewidth}p{0.13\linewidth}p{0.13\linewidth}}
\hline
\begin{tabular}{l}
\textbf{DLT}
\end{tabular} &
\begin{tabular}{l}
\hfil\textbf{Capacity} \\ (transactions \\\hfil per second )
\end{tabular} &
\begin{tabular}{l}
\hfil\textbf{Average} \\ \textbf{Validation Time} \\ \hfil (seconds)
\end{tabular} &
\begin{tabular}{l}
\hfil\textbf{Average} \\ \textbf{Block Size} \\\hfil (bytes)
\end{tabular} &
\begin{tabular}{l}
\hfil \textbf{Size of} \\  \textbf{block header} \\\hfil (bytes)
\end{tabular} &
\begin{tabular}{l}
\textbf{Minimum size} \\ \textbf{of transaction} \\\hfil (bytes)
\end{tabular} 
\\ \hline 
Ethereum &\hfil $1$--$10^2$ &\hfil $1$--$10$ &\hfil $10^3$--$10^4$ &\hfil $508$  &\hfil $110$ \\ 
IOTA &\hfil $10$--$10^3$ &\hfil $10$--$10^2$ &\hfil - &\hfil -  &\hfil $1600^{\star}$ \\ 
Bitcoin &\hfil $1$--$10^1$ &\hfil $10^2$ &\hfil $10^5$--$10^6$  &\hfil $80$ &\hfil $247$  \\ 
Fabric &\hfil $10^3$  &\hfil $10^0$  &\hfil $10^5$--$10^6$  &\hfil $72$ &\hfil $3060^{\star\star}$ \\ \hline 
\end{tabular}
\vspace{0.2cm}
    \begin{tablenotes}
      \small
      \centering
      \item ${}^\star$ of which 1300 bytes are available for data
      \item  ${}^{\star\star}$ of which 960 bytes are available for data
    \end{tablenotes}
\end{table}

We assume a 1000 meter radius cell with a single AP in the center and 100 Class C devices spread uniformly throughout. Spreading factors (SFs) are allocated based on their distance to the AP as in~\cite{sorensen2019analysis}, and we account for both co- and inter-SF interference and the demodulation capabilities of the AP~\cite{sorensen2019analysis}. The UL consists of five 125 kHz UL channels in sub-band G and three in sub-band G1, within the 868 MHz ISM band with a 1\% duty-cycle. UL transmissions are all of the acknowledged type with up to 3 retransmissions. In the DL the block headers are broadcasted using SF 12, that allows reaching devices at the cell edge. Under these conditions, the time to transmit a block header for Ethereum, Bitcoin and Fabric are 24.49 s, 4.11 s and 3.95 s, respectively. This corresponds to a rate limit of, respectively, 0.04, 0.24 and 0.25 block headers per second in a 100\% duty-cycled band.

In the publishing scenario, the devices publish transactions with 50 bytes of payload to the ledger according to a Poisson process with intensity of 1 transaction per week. Once the transaction has been validated by the DLT, a receipt with size 10 bytes is transmitted to the device in the DL, see phase 6 in Fig.~2. Transaction overheads are listed in Table I and the validation times used for Ethereum, Bitcoin, IOTA, and Fabric are 21 seconds, 13 minutes, 88 seconds and 550 milliseconds, respectively. For the case of Fabric, we assume that the endorsement, see phase 0 in Fig.~2, is already available to the device.

\begin{figure}
    \centering
    \includegraphics[width=.7\columnwidth]{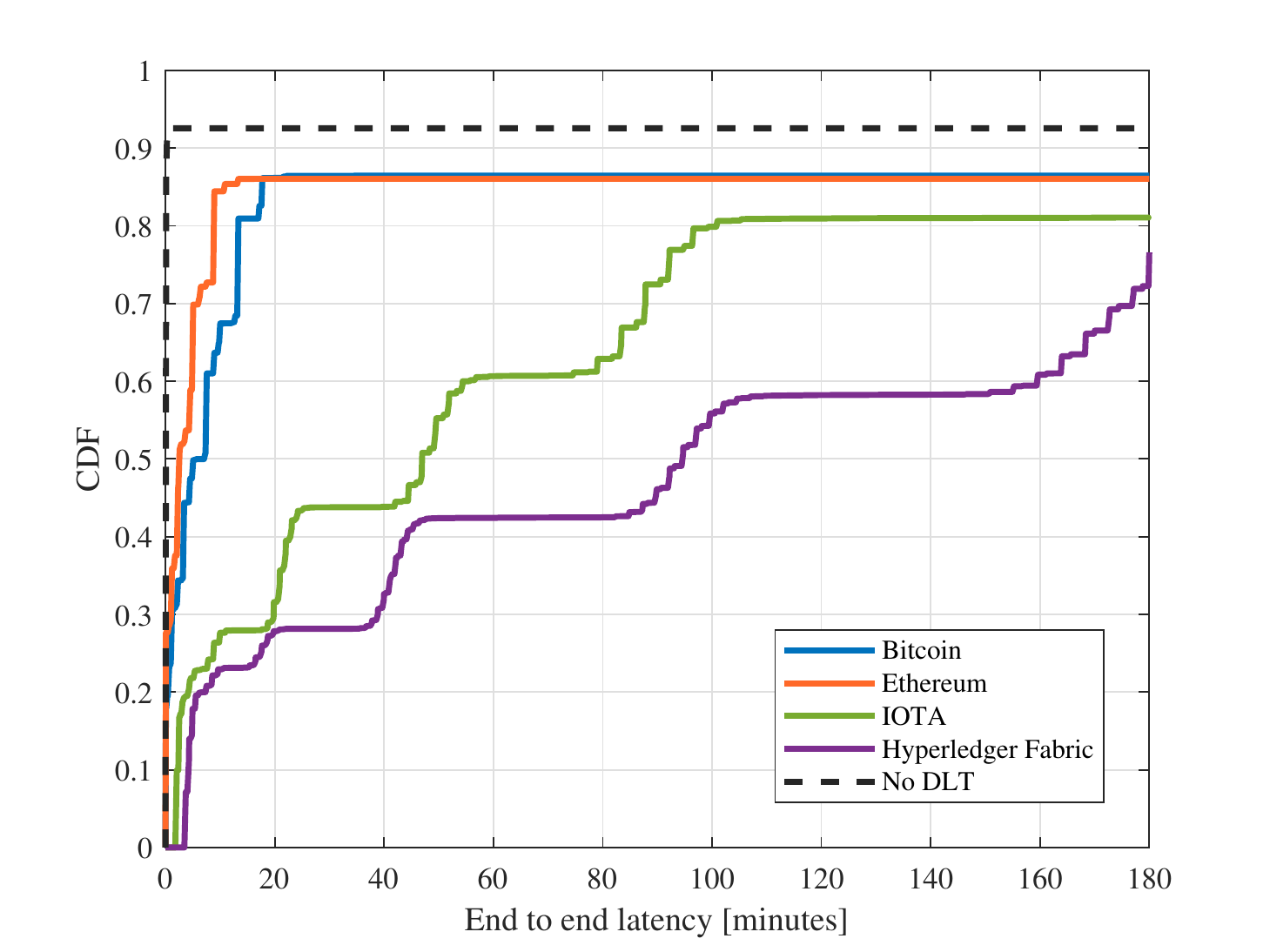}
    \caption{CDF of end-to-end latency for the considered scenario.}
    \label{fig:case1}
\end{figure}

Fig. \ref{fig:case1} shows the Cumulative Density Function (CDF) of the end-to-end latency, defined as the time between a transmission is queued in the device buffer until the receipt is received by the device. In the figure, the horizontal asymptotes are due to transmission failures caused by the unreliable link between the devices and the AP. Obviously, a large transaction size leads to a large number of frames to be transmitted per payload and in turn a higher failure rate, which is mitigated by the acknowledgement-retransmission procedure at the cost of latency. Validation time has a relatively small impact on the latency in this case, but would have a larger impact on unacknowledged operations (Class A).

\subsection{Case study 2: Accountability of operations}
\begin{figure}[t]
\begin{center}
\includegraphics[width=0.85\columnwidth]{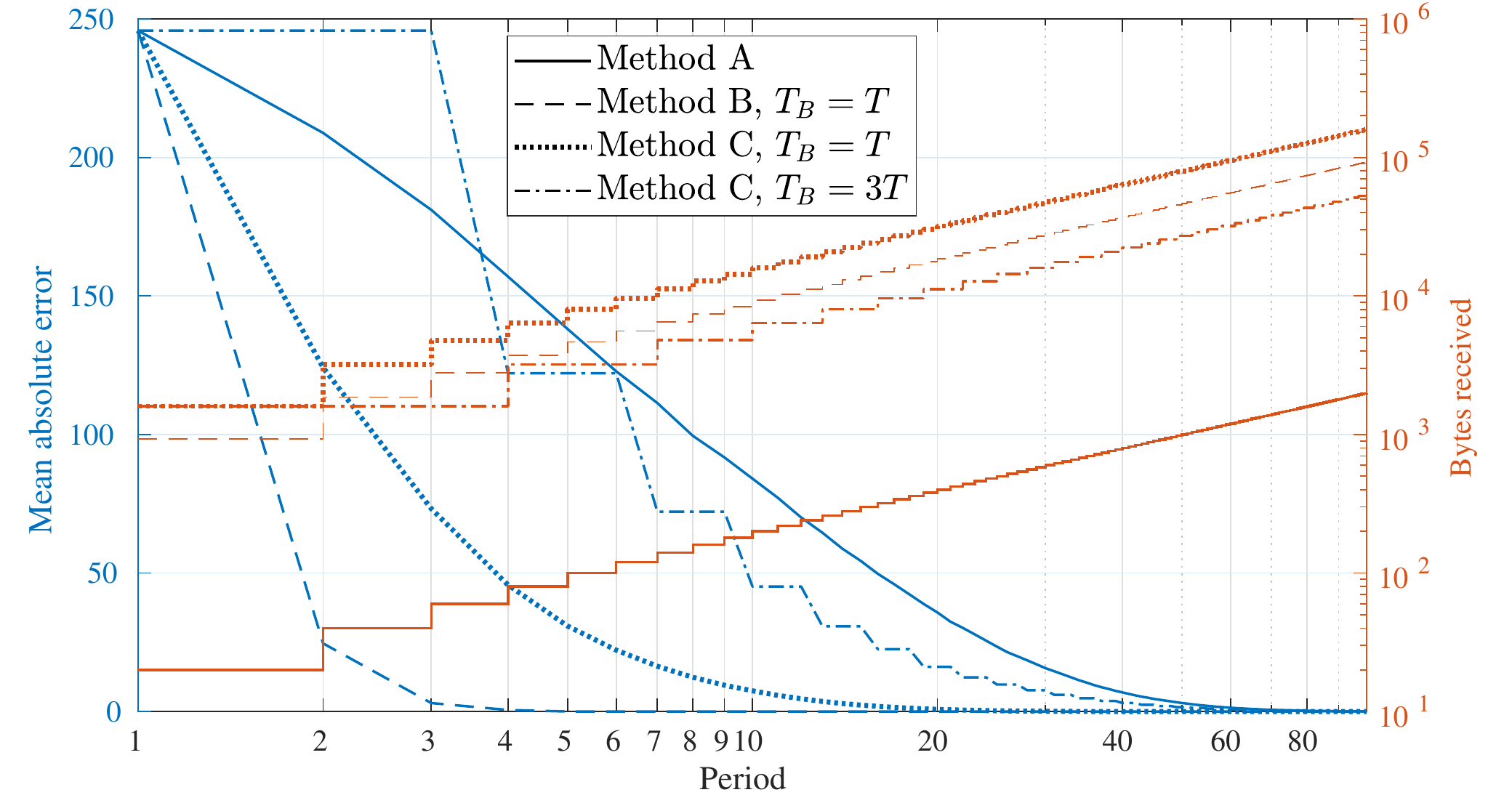}
\end{center}
\caption{Comparison of convergence rate (on the left) and amount of data received by a device (on the right) for the three different methods.}\label{fig:five}
\end{figure}

As a second case study, we consider a scenario with a large deployment of devices that aim to average a value, e.g. a measurement, that is an operation encountered in many applications. Each device only knows a subset of devices, called \emph{neighbors}. The connectivity is provided by IEEE 802.11 APs, see the right part of Fig.~1(b). In the UL each device transmits to the AP to which it is associated with without acknowledgments. In the DL packages are sent from the AP to the device. Both transmissions are assumed to have packet error probability $p$. The UL transmissions happen at a fixed rate, and the averaging procedure is done in transmission periods that occur every $T$ seconds, in which devices exchange their values.

We compare three methods for computing the average: (A) authenticated gossip consensus, (B) smart contract-aided computation and (C) transaction-based computation. In method A, each device sends in each transmission period the locally estimated value to a randomly chosen neighbor device, together with its digital signature. In the same transmission period, the device receives estimates from neighbor devices, and computes the average. After a certain number of exchanges, the devices should converge to the same average value. In method B, all devices connect to a DLT network and send in each transmission period their estimated values to the same smart contract, via transactions. The smart contract is programmed to compute the average upon reception of a new value. All devices subscribe to the updates of this contract and, every time a block is generated, receive the computed average value.
Finally, in method C, %
in each transmission period, each device sends a transaction containing its local value to the DLT network. Furthermore, each device also subscribes to the transactions sent by its neighbor devices. All three methods provide accountability, because the messages sent by devices are always signed. However, the benefit of DLT-based methods (B and C), with respect to A, is that the devices do not need to locally store the signatures of other devices, as they are stored in the ledger.

We simulate an execution of the three methods in a strongly connected network consisting of 1000 devices, each of them knowing as neighbors 5 random devices. The DLT that we use is Ethereum blockchain protocol with block period, $T_B$, equal to the transmission period interval $T$. The packet error probability is $p=0.1$ and equal for all devices, and all messages are sent as a single packet. The estimated values are represented by 32-bit floating points.

Fig.~5 shows that the three methods provide different rates of convergence, and that method B is fastest, as the smart contract acts as a centralized information fusion point.\footnote{Notice that devices are making use of the information received from the DLT as soon as it is received, without waiting for its confirmation. Waiting for confirmation may be required in some applications.} However, we observe a substantial difference in the amount of data exchanged with the APs with each method. Specifically, with method A the device transmits 4 bytes in UL for each transmission period, plus a digital signature, that has size 72 bytes (assuming the use of ECDSA with 256-bit elliptic curve). With methods B and C it transmits 4 bytes, plus the overhead given by the metadata of a transaction, see the last column of Table I. In the DL channel, with method A the device receives a maximum of 20 bytes (5 packets) during a transmission period, while in method B a device receives approximately 931 bytes: the block header (508 bytes), the new state of the smart contract (48 bytes), and the proof that the new state is included in the block (375 bytes)~\cite{danzi2018delay}. Finally, with C the device receives approximately 1600 bytes, given by block header, up to five valid transactions, and the proof that they are included in the block. The figure shows that both convergence and bytes received depend on the ratio between $T$ and $T_B$. The conclusion is that, with DLT-based methods (B and C), the 802.11 APs are subject to a significantly higher load in DL.

\section{Conclusion and outlook}

DLTs are still in an early stage and therefore open to large protocol re-designs.
There are a number of ongoing investigations to improve the core features of DLTs networks in order to reduce their operating costs and offer a lower latency.
This work shows that the design of DLTs should also take into account the capabilities of the edge networks, especially in the context of devices with limited capacity.
We have focused on the communication aspects, introducing the architecture of DLT-based distributed trust networks, and proposed a new classification for lightweight DLT synchronization protocols. Then, we showed potential integration issues by studying the cases of two wireless technologies.
The results clearly show that wireless systems may impede the full functionalities of lightweight synchronization protocols. This should be taken into account by the industry if the support of DLT will become a must-have feature.
On the other hand, we prompt the DLT communities to invest more effort in understanding the limits of the wireless solutions, to design protocols that actually fit in the IoT landscape.

\nocite{*}
\bibliographystyle{IEEEtran}

\end{document}